\documentclass[useAMS,usenatbib]{mn2e}

\usepackage{graphicx} 
\usepackage{amssymb}
\usepackage{journals}
\usepackage[dvips]{color}
\newcommand{\teff}{$T_{\rm eff}$} 
\newcommand{\tcond}{$T_{\rm cond}$} 
\newcommand{\logg}{$\log g$} 
\newcommand{\kms}{km s$^{-1}$}
\newcommand{\vt}{$\xi_t$} 
\newcommand{\fei}{Fe\,{\sc i}}
\newcommand{\feii}{Fe\,{\sc ii}}

\newcommand\simgt{\lower.3ex\hbox{\gtsima}}

\usepackage{placeins}
\usepackage[modulo]{lineno}
\usepackage{hyperref}
\usepackage{rotating}

\title[Chemical compositions of planet hosting stars]{Detailed chemical compositions of planet hosting stars: I. Exploration of possible planet signatures\thanks{The data presented herein were obtained at the W. M. Keck Observatory, which is operated as a scientific partnership among the California Institute of Technology, the University of California and the National Aeronautics and Space Administration. The Observatory was made possible by the generous financial support of the W. M. Keck Foundation.}}
\author[F.\ Liu et al.]
{F.\ Liu,$^{1}$\thanks{E-mail: fanliu@swin.edu.au} 
D.\ Yong,$^{2}$
M.\ Asplund,$^{2}$
H.~S.\ Wang,$^{3}$
L.\ Spina,$^{4,5}$
L.\ Acu\~{n}a,$^{6}$
J.\ Mel\'{e}ndez$^{7}$ \newauthor and 
I.\ Ram\'{i}rez$^{8}$ \\
\\ 
$^{1}$Centre for Astrophysics and Supercomputing, Swinburne University of Technology, Melbourne, VIC 3122, Australia\\
$^{2}$Research School of Astronomy and Astrophysics, Australian National University, Canberra, ACT 2611, Australia\\
$^{3}$Institute for Particle Physics and Astrophysics, ETH Z\"urich, Wolfgang-Pauli-Strasse 27, 8093 Z\"urich, Switzerland\\
$^{4}$Monash Centre for Astrophysics, School of Physics and Astronomy, Monash University, Melbourne, VIC 3800, Australia\\
$^{5}$ARC Centre of Excellence for All Sky Astrophysics in Three Dimensions (ASTRO-3D)\\
$^{6}$Laboratoire d'Astrophysique de Marseille, Aix-Marseille University, 13388 Marseille cedex 13, France\\
$^{7}$Departamento de Astronomia do IAG/USP, Universidade de Sao Paulo, Rua do Matao 1226, Sao Paulo 05508-900, Brasil\\
$^{8}$Tacoma Community College, 6501 South 19th Street, Tacoma, WA 98466, USA}

\begin{document}


\pagerange{\pageref{firstpage}$-$\pageref{lastpage}} \pubyear{2020}

\maketitle

\label{firstpage}

\begin{abstract}
We present a line-by-line differential analysis of a sample of 16 planet hosting stars and 68 comparison stars using high resolution, high signal-to-noise ratio spectra gathered using Keck. We obtained accurate stellar parameters and high-precision relative chemical abundances with average uncertainties in \teff, \logg, [Fe/H] and [X/H] of 15 K, 0.034 [cgs], 0.012 dex and 0.025 dex, respectively. For each planet host, we identify a set of comparison stars and examine the abundance differences (corrected for Galactic chemical evolution effect) as a function of the dust condensation temperature, \tcond, of the individual elements. While we confirm that the Sun exhibits a negative trend between abundance and \tcond, we also confirm that the remaining planet hosts exhibit a variety of abundance $-$ \tcond\ trends with no clear dependence upon age, metallicity or \teff. The diversity in the chemical compositions of planet hosting stars relative to their comparison stars could reflect the range of possible planet-induced effects present in these planet hosts, from the sequestration of rocky material (refractory poor), to the possible ingestion of planets (refractory rich). Other possible explanations include differences in the timescale, efficiency and degree of planet formation or inhomogeneous chemical evolution. Although we do not find an unambiguous chemical signature of planet formation among our sample, the high-precision chemical abundances of the host stars are essential for constraining the composition and structure of their exoplanets. 
\end{abstract}


\section{INTRODUCTION}

The detection and characterisation of exoplanets remains a major focus in modern astronomy. Since the discovery of the first exoplanet around a solar type star \citep{1995Natur.378..355M}, thousands of confirmed exoplanets have been discovered through a raft of methods, the most successful being radial velocity variations and transit photometry. Arguably the greatest surprise has been the enormous diversity in the properties of exoplanets \citep{Winn:2015aa}. 

A star and its planets are believed to arise from the same molecular cloud. The properties of the host star likely influence the proto-planetary disc in which the planets form. Vice versa, accretion of materials from the proto-planetary disc onto the host star may also implant disc/planet signatures into the host stars. Correlations between the properties of planets and the characteristics of their host stars may therefore provide crucial insight into the planet formation process.

Soon after the first exoplanets were discovered, it was recognised that the frequency of hot Jupiters increased rapidly with the overall metallicity of the host star (e.g., \citealt{1997MNRAS.285..403G,2005ApJ...622.1102F}) with iron being the canonical measure of stellar metallicity. The metallicity dependence is typically interpreted as support for the core accretion model of giant planet formation \citep{1996Icar..124...62P} while in the rival disc instability model \citep{1997Sci...276.1836B} no such correlation is expected.

In their pioneering study, \citet[hereafter M09]{Melendez:2009ab} applied a strictly differential line-by-line analysis of high-quality spectra ($R$ = 65,000, signal-to-noise ratio $S/N$ $\sim$ 450) of a sample of solar twins (stars with essentially identical parameters to the Sun). This approach enabled relative chemical abundance measurements to be obtained with unprecedented precision at the $\sim$0.01 dex level (see \citealt{2018A&ARv..26....6N} for an overall review of the method and \citealt{2014ApJ...795...23B} for a discussion of systematic uncertainties). M09 found that the Sun was chemically peculiar when compared to the majority of solar twins (about 15\% of solar twins share the solar chemical composition). A striking correlation was found when plotting the abundance differences (Sun with respect to the average of the solar twins) as a function of the dust condensation temperature (\tcond) of the individual elements.

M09 suggested that this correlation may represent the chemical signature of terrestrial planet formation. While these results have been confirmed and extended to additional planet hosts by some investigators (e.g., \citealt{2009A&A...508L..17R,Ramirez:2010aa,gonzalez:2010aa,schuler:2011aa,Liu:2016aa}), the interpretation remains contentious with other groups finding conflicting results (e.g., \citealt{2010ApJ...720.1592G,2013A&A...552A...6G,Adibekyan:2014aa}). \citet{Nissen:2015aa,Nissen:2016aa} suggested that the dependence of chemical abundance ratios, [X/Fe], on stellar age complicates the planet signature hypothesis of M09. More recently, it has been proposed that the M09 signature possibly arises from the formation of Jupiter analogs (i.e., giant planets at large separations), rather then being a signature of rocky planets \citep{Booth:2020}.

In addition, post-formation accretion of planets onto the host star can also alter its stellar surface abundances. If the host star is polluted after its birth by refractory-rich planetary material, the convective envelope of the star may be enhanced in high-\tcond\ elements (e.g. \citealp{Pinsonneault:2001apjl}). Such a process can produce the \tcond-dependent trend showing enrichment of chemical abundances for refractory elements in the planet host star. This has been reported in a few spectroscopic studies of wide binaries (e.g., \citealp{Oh:2018apj}, \citealp{Liu:2018aa}, \citealp{Ramirez:2019mnras} and \citealp{Nagar:2019apjl}). Recent studies also found evidences in open clusters \citep{Spina:2015aa,Spina:2018apj}. For example, in \citet{Spina:2015aa} they found that the anomalous star is more metal-rich than the average metallicity of the cluster stars, meaning that such a star has been polluted and not vice versa.

Other hypothesis that has been explored regarding the trend with condensation temperature found by M09, is dust cleansing by massive stars from the same cluster where the Sun was born \citep{Gustafsson:2018a}, or even by radiation of the proto-Sun itself \citep{Gustafsson:2018b}. 

To further explore any potential stellar chemical signature of planet formation, we present a differential chemical abundance analysis of 16 planet hosting stars. The sample selection, observations, data reduction and analysis are described in Section 2. The results and discussion are in Sections 3 and 4, respectively. We summarise and conclude in Section 5.

\section{SAMPLE SELECTION, OBSERVATIONS, DATA REDUCTION AND ANALYSIS}

Potential planet hosts were identified by querying the \url{exoplanets.org} website \citep{2014PASP..126..827H}. Planets with $M$ $<$ 0.1 $M_{\rm Jupiter}$ orbiting sufficiently bright stars ($V$ $<$ 12.5) with effective temperatures well suited for a differential abundance analysis (5400 $<$ \teff\ $<$ 6400 K) were selected.

A set of comparison stars (objects not known at that time to host planets) was identified from various literature sources including \citet{Ramirez:2005ab}, \citet{Bensby:2014aa} and \citet{Reddy:2003aa}. The primary criterion was to select bright stars with similar \teff, \logg\ and [Fe/H] to the planet hosts. Since the goal of this study was to identify whether or not planet hosting stars are chemically peculiar, it is important to minimise the abundance errors by ensuring that the stellar parameters of the planet hosts and comparison stars are as similar as possible. This selection process, however, suffered from the heterogeneous nature of the sample (i.e., there are systematic offsets in the stellar parameters between the various publications). To mitigate this potential problem, Gaia data \citep{2016A&A...595A...1G,2018A&A...616A...1G} were also employed to ensure that the comparison stars occupied similar locations in the $M_G$ versus $BP-RP$ colour magnitude diagram.

\begin{figure}
\centering
\includegraphics[width=.99\hsize]{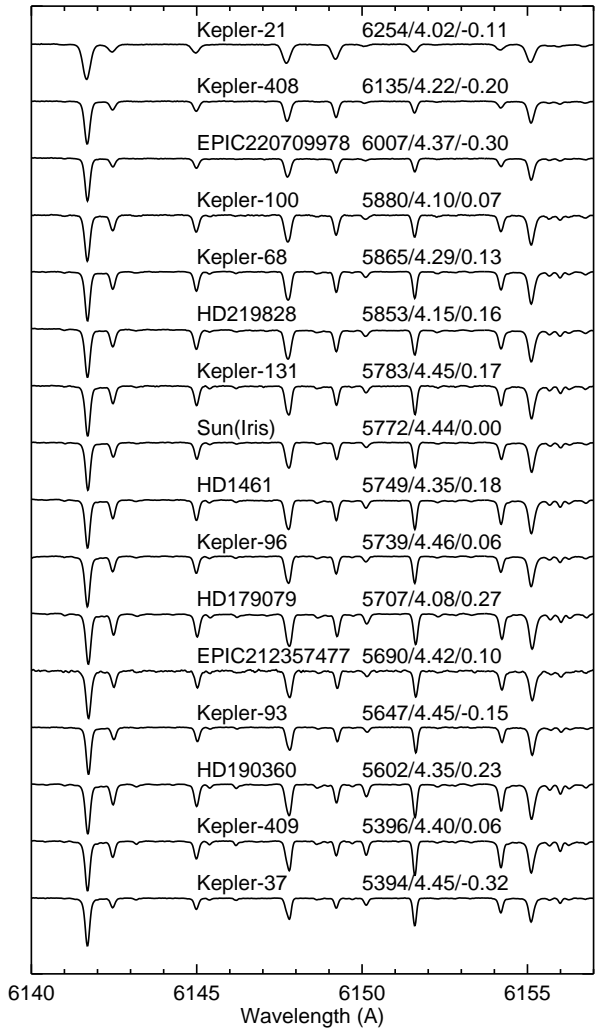}
\caption{A portion of the spectra for the 16 planet hosts. The stellar parameters (\teff, \logg\ and [Fe/H]) are included in the figure and the stars are ordered by \teff\ from hot (top) to cool (bottom).
\label{fig:spectra}}
\end{figure}

A total of 16 planet hosts and 68 comparison stars were observed using the High Resolution Echelle Spectrometer (HIRES; \citealt{1994SPIE.2198..362V}) on the 10m Keck I telescope on 2016-05-18, 2016-08-13 and 2018-07-04. Following \citet{2014MNRAS.442L..51L}, we used the B2 slit with a width of 0\farcs574 which provided a spectral resolution of $R$ = 67,000. For all objects, we aimed to achieve $S/N$ = 300 per pixel near 6000\AA. A higher $S/N$ spectrum was obtained for the asteroid Iris which we use hereafter as a solar spectrum. The wavelength coverage is nearly complete from 3800\AA\ to 8000\AA. The MAuna Kea Echelle Extraction (MAKEE) data reduction package at the Keck Observatory was used to reduce the raw HIRES spectra. Individual spectra were co-added and normalised using {\sc iraf}\footnote{{\sc iraf} is distributed by the National Optical Astronomy Observatories, which are operated by the Association of Universities for Research in Astronomy, Inc., under cooperative agreement with the National Science Foundation.}. Example of a portion of the spectra are shown in Figure \ref{fig:spectra}.

Equivalent widths were measured in each star for a set of lines taken from \citet{2014MNRAS.442L..51L}. The line list and equivalent width measurements are presented in Table A1. Following M09 and \citet{2012A&A...543A..29M}, the effective temperature (\teff) was obtained by forcing excitation balance for \fei\ lines based on a line-by-line differential analysis with respect to the Sun. The surface gravity (\logg) was set when ionization balance between \fei\ and \feii\ was achieved. The microturbulent velocity (\vt) was obtained when there was no trend between the abundance from \fei\ lines and the reduced equivalent width. Following \citet{2014A&A...572A..48R}, we used the qoyllur-quipu\footnote{Qoyllur-quipu or q$^2$ is a PYTHON package available online at \url{https://github.com/astroChasqui/q2}.} code to perform the differential abundance analysis and to compute the uncertainties. The adopted parameters for the Sun are fixed as: T$_{\rm eff} = 5772$\,K, $\log g$ = 4.44 [cm\,s$^{-2}$], $\xi_{\rm t}$ = 1.00 km\,s$^{-1}$, and [Fe/H] = 0.00 dex. Stellar ages were also computed using q$^2$ by fitting Yonsei-Yale isochrones \citep{Demarque:2004apjs}, based on the Gaia DR2 distances as presented in \citet{2018AJ....156...58B}, \teff\ and the metallicity, [Fe/H], from this study. As noted in \citet{Lin:2020mnras}, multiple factors can lead to high and unrealistic ages from isochrone fitting. We note that the oldest stars in our sample have ages of 15.1 Gyr, but that none of the objects older than 10 Gyr are used in the abundance comparisons with the planet hosts.

The uncertainties for the stellar parameters \teff, \logg\ and \vt\ were computed using the approach in \citet{Epstein:2010apj} and \citet{Bensby:2014aa}. For [Fe/H], the uncertainty was obtained by propagating the errors for the stellar parameters noted above then adding them in quadrature (which conservatively assumes that they are uncorrelated) and including the standard error of the mean from line-to-line scatter. For the uncertainties in abundance ratios [X/H], we adopt the same approach as for [Fe/H].

The differential stellar parameters and their uncertainties are presented in Table \ref{tab:param}. The abundances and their uncertainties are given in Table \ref{tab:abund}. With the exception of O, all abundances were derived under the approximation of local thermodynamic equilibrium (LTE). For the O abundances which were derived from the 777nm triplets, non-LTE corrections from \citet{ramirez:2007aa} were applied.

\begin{figure}
\centering
\includegraphics[width=.99\hsize]{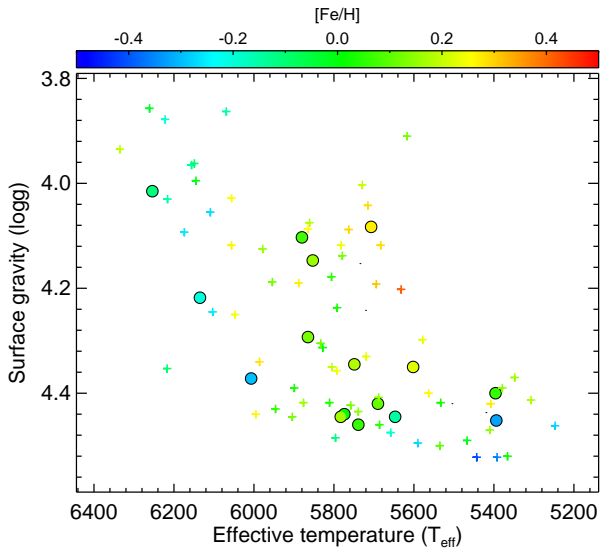}
\caption{Effective temperature (\teff) versus surface gravity (\logg). Planet hosts are shown as circles while the comparison stars are crosses; all objects are coloured by metallicity. There are sufficient comparison stars locate around each planet host.
\label{fig:cmd}}
\end{figure}

In Figure \ref{fig:cmd}, we show the locations of the stars in the \teff\ versus \logg\ plane with the data colour coded by the metallicity. All stars lie on the main sequence or main sequence turn-off and have metallicities in the range $-$0.32 $\le$ [Fe/H] $\le$ $+$0.27. We note there are $\sim$ 5 - 12 comparison stars locate around each planet host. The stellar parameters have small uncertainties due to the high quality spectra and differential nature of the analysis. The average uncertainties for \teff, \logg, \vt\ and [Fe/H] are 15 K, 0.034 [cgs], 0.035 \kms\ and 0.012 dex, respectively. We note that the average uncertainties in stellar parameters for the planet hosts are essentially identical to the corresponding values for the comparison stars.

\begin{table*}
 \centering
 \begin{minipage}{190mm}
  \caption{Stellar parameters and ages for the program stars.} 
  \label{tab:param} 
  \begin{tabular}{@{}lccccccccc@{}}
  \hline
        Name & 
	RA (J2000) &
	Dec (J2000) &
	\teff\ &
	\logg\ & 
	\vt\ &
	[Fe/H] &
	Age &
	Age$^a$ & 
	Age$^b$ \\ 
        & 
	&
	&
	(K) &
	[cgs] & 
	(\kms) &
	(dex) &
	(Gyr) &
	(Gyr) &
	(Gyr) \\ 
\hline 
  \hline
     \multicolumn{10}{c}{Planet hosts}   \\ 
\\ 
Sun (Iris)     &       \ldots &          \ldots & 5774              &  4.440              &  1.00              &    0.000              &   4.6 &       &       \\
HD 1461        &  00 18 42.30 &  $-$08 03 13.03 & 5749 $\pm$     10 &  4.345 $\pm$  0.029 &  1.05 $\pm$   0.03 &    0.180 $\pm$  0.010 &   5.3 &   4.8 &   5.9 \\
EPIC 220709978 &  01 05 51.00 &    +11 45 13.38 & 6007 $\pm$     19 &  4.372 $\pm$  0.049 &  1.32 $\pm$   0.05 & $-$0.296 $\pm$  0.014 &   6.0 &   5.3 &   6.7 \\
EPIC 212357477 &  13 28 03.88 &  $-$15 56 16.70 & 5690 $\pm$     16 &  4.420 $\pm$  0.049 &  1.01 $\pm$   0.05 &    0.100 $\pm$  0.017 &   3.0 &   1.6 &   4.5 \\
Kepler-37      &  18 56 14.22 &    +44 31 06.14 & 5394 $\pm$      8 &  4.452 $\pm$  0.026 &  0.77 $\pm$   0.03 & $-$0.322 $\pm$  0.008 &   7.9 &   5.8 &  10.1 \\
Kepler-408     &  18 59 08.69 &    +48 25 23.74 & 6135 $\pm$     19 &  4.218 $\pm$  0.045 &  1.45 $\pm$   0.04 & $-$0.201 $\pm$  0.014 &   4.2 &   3.8 &   4.8 \\
Kepler-21      &  19 09 26.87 &    +38 42 50.89 & 6254 $\pm$     25 &  4.015 $\pm$  0.044 &  1.74 $\pm$   0.06 & $-$0.110 $\pm$  0.016 &   3.8 &   3.2 &   4.1 \\
HD 179079      &  19 11 09.68 &  $-$02 38 19.57 & 5707 $\pm$     22 &  4.083 $\pm$  0.042 &  1.17 $\pm$   0.04 &    0.272 $\pm$  0.018 &   6.5 &   6.1 &   6.8 \\
Kepler-131     &  19 14 07.41 &    +40 56 32.54 & 5783 $\pm$     10 &  4.445 $\pm$  0.026 &  1.03 $\pm$   0.03 &    0.166 $\pm$  0.011 &   1.4 &   0.6 &   2.6 \\
Kepler-68      &  19 24 07.75 &    +49 02 24.76 & 5865 $\pm$      6 &  4.293 $\pm$  0.016 &  1.13 $\pm$   0.01 &    0.126 $\pm$  0.005 &   5.3 &   4.9 &   5.9 \\
Kepler-93      &  19 25 40.35 &    +38 40 20.32 & 5647 $\pm$      6 &  4.445 $\pm$  0.020 &  0.92 $\pm$   0.02 & $-$0.146 $\pm$  0.006 &   6.9 &   5.6 &   7.8 \\
Kepler-100     &  19 25 32.67 &    +41 59 24.50 & 5880 $\pm$     18 &  4.103 $\pm$  0.035 &  1.28 $\pm$   0.03 &    0.070 $\pm$  0.014 &   6.6 &   6.2 &   6.9 \\
Kepler-409     &  19 34 43.00 &    +46 51 09.84 & 5396 $\pm$     12 &  4.400 $\pm$  0.035 &  0.82 $\pm$   0.04 &    0.059 $\pm$  0.012 &   8.9 &   6.9 &  10.2 \\
Kepler-96      &  19 48 16.75 &    +40 31 30.73 & 5739 $\pm$      4 &  4.460 $\pm$  0.014 &  0.95 $\pm$   0.01 &    0.059 $\pm$  0.004 &   2.6 &   1.2 &   3.2 \\
HD 190360      &  20 03 38.21 &    +29 53 40.34 & 5602 $\pm$     17 &  4.350 $\pm$  0.042 &  1.04 $\pm$   0.05 &    0.228 $\pm$  0.017 &   7.6 &   6.7 &   8.9 \\
HD 219828      &  23 18 46.72 &    +18 38 44.70 & 5853 $\pm$     17 &  4.147 $\pm$  0.034 &  1.27 $\pm$   0.03 &    0.160 $\pm$  0.013 &   6.0 &   5.4 &   6.3 \\
\\ 
     \multicolumn{10}{c}{Comparison stars}   \\ 
\\ 
HD 166         &  00 06 37.23 &    +29 01 14.65 & 5535 $\pm$     13 &  4.500 $\pm$  0.031 &  1.18 $\pm$   0.04 &    0.099 $\pm$  0.013 &   1.7 &   0.6 &   3.4 \\
HD 4813        &  00 50 07.34 &  $-$10 38 43.14 & 6217 $\pm$     19 &  4.353 $\pm$  0.055 &  1.34 $\pm$   0.05 & $-$0.095 $\pm$  0.013 &   2.7 &   2.2 &   3.1 \\
HD 4915        &  00 51 11.11 &  $-$05 02 23.24 & 5658 $\pm$      8 &  4.475 $\pm$  0.026 &  0.94 $\pm$   0.02 & $-$0.192 $\pm$  0.007 &   5.0 &   3.5 &   6.3 \\
HD 6250        &  01 04 16.35 &    +54 12 14.31 & 6216 $\pm$     22 &  4.030 $\pm$  0.039 &  1.72 $\pm$   0.05 & $-$0.150 $\pm$  0.015 &   4.0 &   3.3 &   4.3 \\
HD 7438        &  01 14 22.55 &  $-$07 54 35.04 & 5247 $\pm$     13 &  4.462 $\pm$  0.037 &  0.74 $\pm$   0.05 & $-$0.278 $\pm$  0.011 &  10.3 &   7.4 &  12.7 \\
 \hline
\end{tabular}
\end{minipage}
{\raggedright $^a$ Lower limit \\ $^b$ Upper limit \\ \par}
\vspace{1ex}
This table is published in its entirety in the electronic edition of the paper. A portion is shown here for guidance regarding its content.
\end{table*}

\begin{table*}
 \centering
 \begin{minipage}{190mm}
  \caption{Relative abundances for the program stars.} 
  \label{tab:abund} 
  \begin{tabular}{@{}lrcrcrcrcrcc@{}}
  \hline
Name &
[C/H] &
$\sigma$[C/H] &
[O/H] &
$\sigma$[O/H] &
[Na/H] &
$\sigma$[Na/H] &
[Mg/H] &
$\sigma$[Mg/H] &
[Al/H] &
$\sigma$[Al/H] &
\ldots \\
 \hline 
   \hline
      \multicolumn{12}{c}{Planet hosts}   \\ 
 \\ 
Sun (Iris)     &     0.000 & \ldots &      0.000 & \ldots &      0.000 & \ldots &      0.000 & \ldots &      0.000 & \ldots & \ldots \\ 
HD 1461         &     0.215 &  0.033 &      0.126 &  0.018 &      0.291 &  0.030 &      0.173 &  0.009 &      0.260 &  0.029 & \ldots \\ 
EPIC220709978  &  $-$0.191 &  0.122 &      $-$0.254 &  0.021 &      $-$0.261 &  0.028 &      $-$0.259 &  0.014 &      $-$0.316 &  0.038 & \ldots \\ 
EPIC212357477  &     0.167 &  0.066 &      0.092 &  0.027 &      0.091 &  0.027 &      0.059 &  0.016 &      0.132 &  0.010 & \ldots \\ 
Kepler-37      &  $-$0.284 &  0.030 &      $-$0.201 &  0.011 &      $-$0.340 &  0.018 &      $-$0.199 &  0.009 &      $-$0.150 &  0.102 & \ldots \\ 
Kepler-408     &  $-$0.178 &  0.028 &      $-$0.205 &  0.024 &      $-$0.232 &  0.039 &      $-$0.150 &  0.013 &      $-$0.220 &  0.022 & \ldots \\ 
Kepler-21      &  $-$0.091 &  0.043 &      $-$0.069  & 0.028 &      $-$0.065 &  0.045 &      $-$0.111 &  0.015 &      $-$0.237 &  0.009 & \ldots \\ 
HD 179079       &     0.381 &  0.046 &      0.205 &  0.037 &      0.382 &  0.023 &      0.272 &  0.017 &      0.378 &  0.022 & \ldots \\ 
Kepler-131     &     0.093 &  0.070 &      0.072 &  0.015 &      0.125 &  0.023 &      0.129 &  0.010 &      0.200 &  0.014 & \ldots \\ 
Kepler-68      &     0.053 &  0.015 &      0.063 &  0.010 &      0.139 &  0.016 &      0.130 &  0.005 &      0.313 &  0.126 & \ldots \\ 
Kepler-93      &  $-$0.163 &  0.040 &      $-$0.097 &  0.010 &      $-$0.161 &  0.017 &      $-$0.095 &  0.006 &      $-$0.039 &  0.038 & \ldots \\ 
Kepler-100     &     0.171 &  0.020 &      0.078 &  0.026 &      0.151 &  0.019 &      0.091 &  0.013 &      0.152 &  0.028 & \ldots \\ 
Kepler-409     &     0.085 &  0.015 &      \ldots  & \ldots &    0.094 &  0.048 &      0.096 &  0.014 &      0.215 &  0.050 & \ldots \\ 
Kepler-96      &  $-$0.045 &  0.021 &      $-$0.007 &  0.007 &      $-$0.046 &  0.007 &      0.017 &  0.004 &      0.010 &  0.028 & \ldots \\ 
HD 190360       &     0.386 &  0.034 &      0.240 &  0.021 &      0.270 &  0.037 &      0.346 &  0.018 &      0.455 &  0.071 & \ldots \\ 
HD 219828       &     0.172 &  0.047 &      0.136 &  0.025 &      0.271 &  0.022 &      0.144 &  0.014 &      0.260 &  0.025 & \ldots \\ 
\\ 
     \multicolumn{12}{c}{Comparison stars}   \\ 
\\ 
HD 166          &     0.081 &  0.043 &      0.070 &  0.017 &      0.069 &  0.027 &      0.064 &  0.015 &      0.103 &  0.017 & \ldots \\ 
HD 4813         &  $-$0.055 &  0.096 &      $-$0.089 &  0.024 &      $-$0.129 &  0.032 &      $-$0.120 &  0.013 &      $-$0.230 &  0.032 & \ldots \\ 
HD 4915         &  $-$0.005 &  0.008 &      $-$0.140 &  0.010 &      $-$0.234 &  0.015 &      $-$0.150 &  0.007 &      $-$0.175 &  0.035 & \ldots \\ 
HD 6250         &  $-$0.102 &  0.026 &      $-$0.129 &  0.031 &      $-$0.114 &  0.043 &      $-$0.145 &  0.013 &      $-$0.209 &  0.021 & \ldots \\ 
HD 7438         &  $-$0.087 &  0.013 &      $-$0.167 &  0.017 &      $-$0.320 &  0.027 &      $-$0.189 &  0.014 &      $-$0.207 &  0.016 & \ldots \\ 
 \hline
\end{tabular}
\end{minipage}
This table is published in its entirety in the electronic edition of the paper. A portion is shown here for guidance regarding its content.
\end{table*}

\begin{figure}
\centering
\includegraphics[width=.99\hsize]{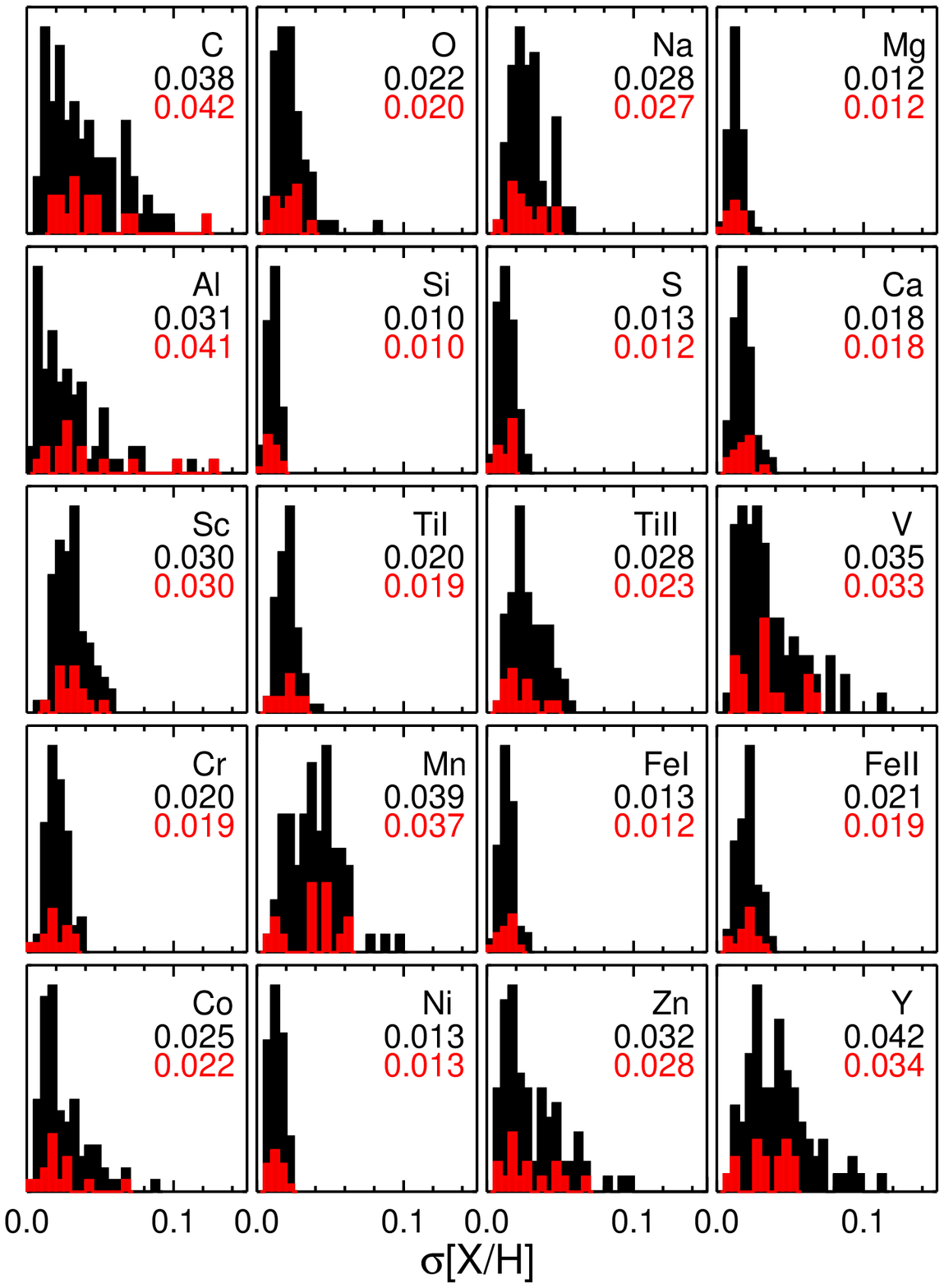}
\caption{The error distribution for each element, [X/H], for all stars (black histogram) and planet hosts (red histogram). The average error is written in each panel.
\label{fig:err}}
\end{figure}

The distribution of the abundance uncertainties for each element, $\sigma$[X/H], is presented in Figure \ref{fig:err}. The first point to note is that the average abundance uncertainties are small, in the range 0.010 dex (for Si) to 0.042 dex (for C). Such high-precision relative chemical abundance measurements can be attributed to the small uncertainties in stellar parameters which in turn are due to the differential analysis of high quality spectra. The second point to note is that, modulo small number statistics, the average abundance error for a given element is the same for the planet host population and for the comparison stars.

\section{RESULTS} 

\begin{figure*}
\centering
\includegraphics[width=.75\hsize]{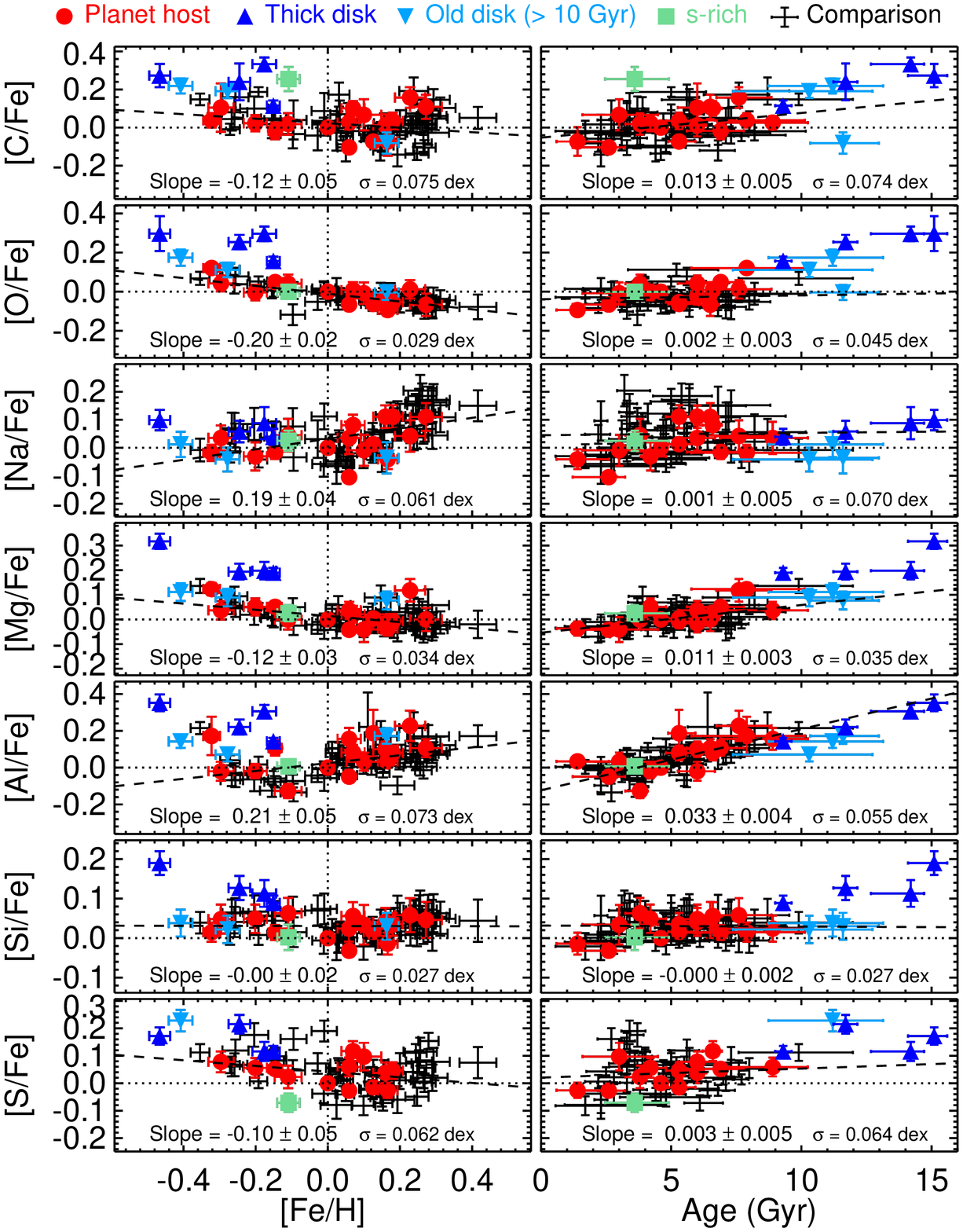}
\caption{[X/Fe] versus [Fe/H] (left panels) and ages (right panels) for the first seven elements. The best fitting slope to the comparison stars is overplotted. The slope and dispersion about the fit is written in each panel. 
\label{fig:xfe_feh_age_a}}
\end{figure*}

\begin{figure*}
\centering
\includegraphics[width=.75\hsize]{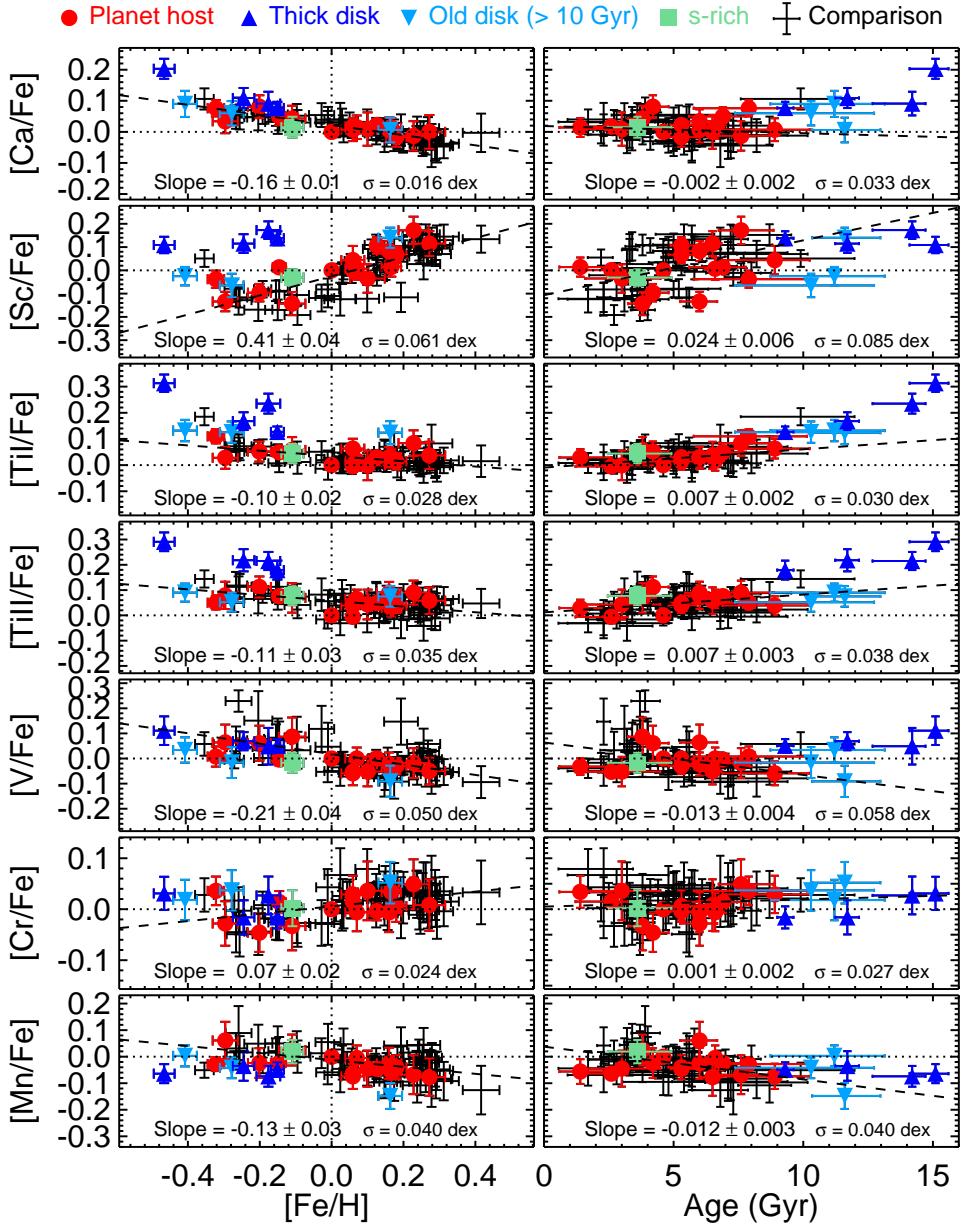}
\caption{Same as Figure \ref{fig:xfe_feh_age_a} but for the next seven elements. 
\label{fig:xfe_feh_age_b}}
\end{figure*}

\begin{figure*}
\centering
\includegraphics[width=.75\hsize]{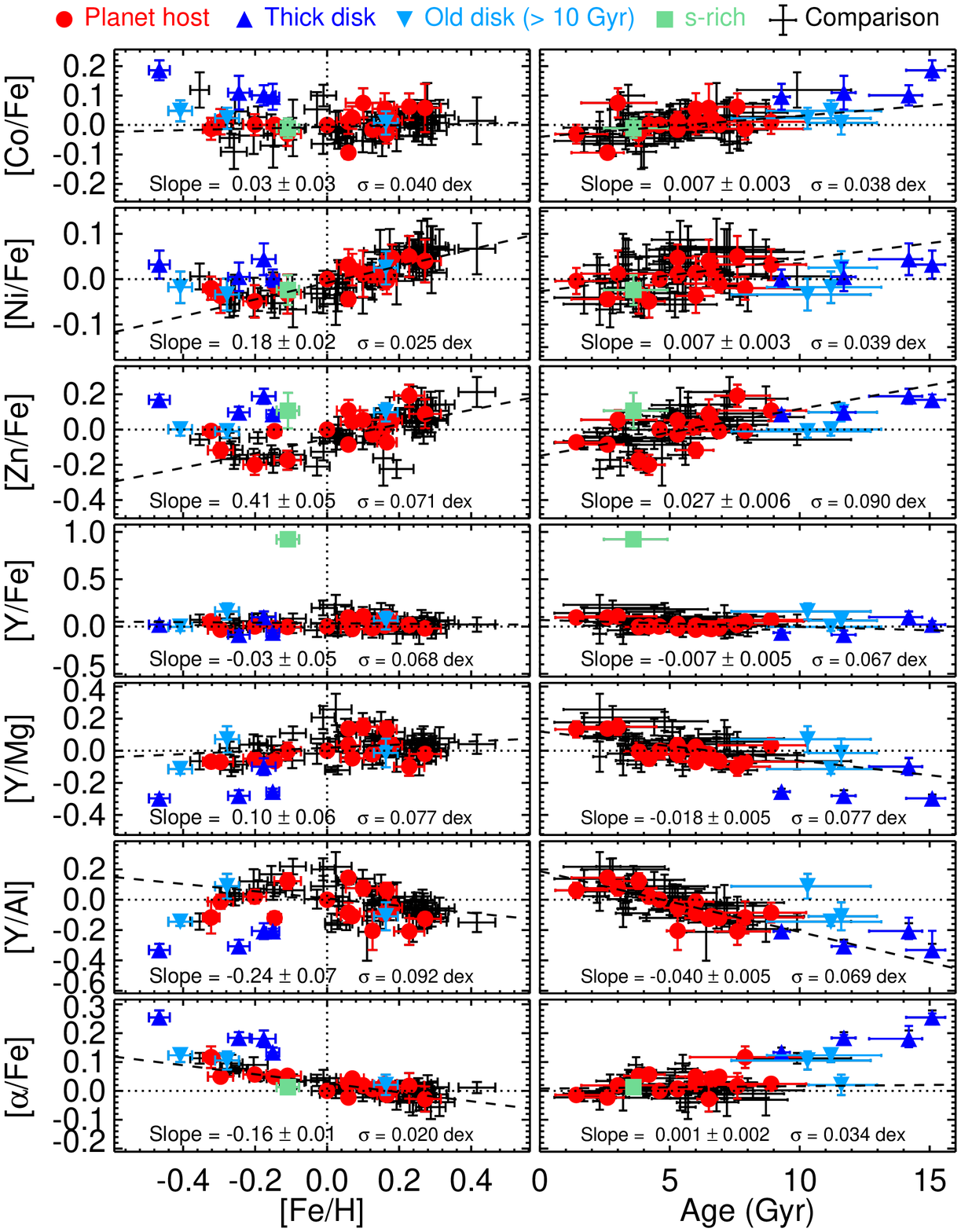}
\caption{Same as Figure \ref{fig:xfe_feh_age_a} but for the final seven element ratios. 
\label{fig:xfe_feh_age_c}}
\end{figure*}

We now examine the trends between abundance ratios [X/Fe] versus [Fe/H] (left panels) in Figures \ref{fig:xfe_feh_age_a} to \ref{fig:xfe_feh_age_c}. In all panels, we present the linear fit to the comparison stars, i.e., we exclude from the fit the planet hosts, thick disc objects, stars with ages $\ge$ 10 Gyr and HD~138004 with [Y/Fe] = $+$0.92 which we hereafter refer to as a $s$-process rich object.

In Table \ref{tab:gce_feh}, we present the coefficients $a$ and $b$ for the fit to the relation [X/Fe] = $a$ $\times$ [Fe/H] + $b$. The abundance trends [X/Fe] versus [Fe/H] are well-defined, and the dispersion about the linear fit is small, ranging from 0.016 dex (for Ca) to 0.075 dex (for C) with an average value of 0.046 dex.

Overall, our abundance measurements are of high precision due to the high quality spectra and differential abundance analysis. We note, however, that the analyses by \citet{Spina:2018mnras} and \citet{Bedell:2018aa} achieved even better abundance precision most likely because their sample spanned a considerably smaller range in stellar parameters $\Delta$\teff\ = 283 K, $\Delta$\logg\ = 0.445 [cgs] and $\Delta$[Fe/H] = 0.285 dex while the corresponding values for this study are 1088 K, 0.665 [cgs] and 0.882 dex, respectively.

Examination of the [$\alpha$/Fe] versus [Fe/H] in Figure \ref{fig:xfe_feh_age_c} indicates that there are four stars which lie above the general trend. We refer to those objects as thick disc stars as they exhibit abundance trends in accordance with that stellar population (e.g., \citealt{Venn:2004aa}), and we note that their ages are consistent with an older population; HD~11505 (11.7 Gyr), HIP~14241 (15.1 Gyr), HD~136274 (14.2 Gyr) and HD~187923 (9.3 Gyr). Additionally, there are three stars with ages greater than 10 Gyr. While these objects exhibit thin-disc like [$\alpha$/Fe] ratios, we highlight these stars in various figures and refer to them as "old disc" objects; HD~7438 (10.3 Gyr), HD~164922 (11.6 Gyr) and HD~172310 (11.2 Gyr).

As recognised by \citet{Nissen:2015aa}, \citet{Liu:2016aa}, \citet{Spina:2016aa}, \citet{Bedell:2018aa} and others, it is important to consider Galactic chemical evolution (GCE) effects when comparing stars with different ages even if they have similar \teff\ and [Fe/H]. To that end, we present the abundance trends [X/Fe] versus age (right panels) in Figures \ref{fig:xfe_feh_age_a} to \ref{fig:xfe_feh_age_c}. As before, we include the linear fit to the comparison stars excluding the planet hosts, thick disc, old disc and $s$-rich objects. Table \ref{tab:gce_age} includes the coefficients $a$ and $b$ for the relation [X/Fe] = $a$ $\times$ Age + $b$. We confirm previous results that the [Y/Al] ratio exhibits a steep dependence upon stellar age, $-$0.040 dex Gyr$^{-1}$. For comparison, \citet{Nissen:2016aa} and \citet{Spina:2018mnras} find slopes of $-$0.0427 and $-$0.051 dex Gyr$^{-1}$, respectively. Based on the coefficients presented in Table \ref{tab:gce_age}, the [Y/Zn] ratio would also exhibit a steep dependence on age, $-$0.034 dex Gyr$^{-1}$ (see also \citealp{Jofre:2020aa} for other potential chemical clocks).

\begin{table}
 \centering
 \begin{minipage}{85mm}
  \caption{Results of the linear fitting of [X/Fe] versus [Fe/H] for the
comparison stars excluding thick disc objects and stars with ages $\ge$ 10 Gyr. The
coefficients are [X/Fe] = $a$ $\times$ [Fe/H] + $b$.}
  \label{tab:gce_feh} 
  \begin{tabular}{@{}lrr@{}}
  \hline
Name &
$a$ & 
$b$ \\ 
&
(dex$^{-1}$) &  
(dex) \\ 
 \hline 
   \hline
$\rm [C/Fe]$    & $-$0.118 $\pm$ 0.055 &    0.023 $\pm$ 0.011 \\  
$\rm [O/Fe]$    & $-$0.200 $\pm$ 0.023 & $-$0.010 $\pm$ 0.004 \\  
$\rm [Na/Fe]$   &    0.188 $\pm$ 0.045 &    0.032 $\pm$ 0.009 \\  
$\rm [Mg/Fe]$   & $-$0.125 $\pm$ 0.025 &    0.014 $\pm$ 0.005 \\  
$\rm [Al/Fe]$   &    0.214 $\pm$ 0.053 &    0.025 $\pm$ 0.011 \\  
$\rm [Si/Fe]$   & $-$0.001 $\pm$ 0.019 &    0.031 $\pm$ 0.004 \\  
$\rm [S/Fe]$    & $-$0.103 $\pm$ 0.048 &    0.042 $\pm$ 0.009 \\  
$\rm [Ca/Fe]$   & $-$0.163 $\pm$ 0.011 &    0.022 $\pm$ 0.002 \\  
$\rm [Sc/Fe]$   &    0.411 $\pm$ 0.044 & $-$0.025 $\pm$ 0.009 \\  
$\rm [TiI/Fe]$  & $-$0.101 $\pm$ 0.020 &    0.035 $\pm$ 0.004 \\  
$\rm [TiII/Fe]$ & $-$0.113 $\pm$ 0.025 &    0.058 $\pm$ 0.005 \\  
$\rm [V/Fe]$    & $-$0.207 $\pm$ 0.037 &    0.018 $\pm$ 0.007 \\  
$\rm [Cr/Fe]$   &    0.072 $\pm$ 0.017 &    0.006 $\pm$ 0.003 \\  
$\rm [Mn/Fe]$   & $-$0.131 $\pm$ 0.029 & $-$0.012 $\pm$ 0.006 \\  
$\rm [Co/Fe]$   &    0.027 $\pm$ 0.029 & $-$0.007 $\pm$ 0.006 \\  
$\rm [Ni/Fe]$   &    0.184 $\pm$ 0.018 & $-$0.008 $\pm$ 0.004 \\  
$\rm [Zn/Fe]$   &    0.411 $\pm$ 0.052 & $-$0.052 $\pm$ 0.010 \\  
$\rm [Y/Fe]$    & $-$0.028 $\pm$ 0.050 &    0.035 $\pm$ 0.010 \\  
 \hline
\end{tabular}
\end{minipage}
\end{table}

\begin{table}
 \centering
 \begin{minipage}{85mm}
  \caption{Results of the linear fitting of [X/Fe] versus ages for the comparison stars excluding thick disc objects and stars with ages $\ge$ 10 Gyr. The coefficients are [X/Fe] = $a$ $\times$ Age + $b$.}
  \label{tab:gce_age} 
  \begin{tabular}{@{}lrr@{}}
  \hline
Name &
$a$ & 
$b$ \\ 
&
(Gyr$^{-1}$) &  
(dex) \\ 
 \hline 
   \hline
$\rm [C/Fe]   $ &    0.013 $\pm$ 0.005 & $-$0.053 $\pm$ 0.028 \\ 
$\rm [O/Fe]   $ &    0.002 $\pm$ 0.003 & $-$0.039 $\pm$ 0.018 \\ 
$\rm [Na/Fe]  $ &    0.001 $\pm$ 0.005 &    0.045 $\pm$ 0.026 \\ 
$\rm [Mg/Fe]  $ &    0.011 $\pm$ 0.003 & $-$0.054 $\pm$ 0.013 \\ 
$\rm [Al/Fe]  $ &    0.033 $\pm$ 0.004 & $-$0.123 $\pm$ 0.021 \\ 
$\rm [Si/Fe]  $ & $-$0.000 $\pm$ 0.002 &    0.032 $\pm$ 0.010 \\ 
$\rm [S/Fe]   $ &    0.003 $\pm$ 0.005 &    0.020 $\pm$ 0.026 \\ 
$\rm [Ca/Fe]  $ & $-$0.002 $\pm$ 0.002 &    0.019 $\pm$ 0.013 \\ 
$\rm [Sc/Fe]  $ &    0.024 $\pm$ 0.006 & $-$0.107 $\pm$ 0.032 \\ 
$\rm [TiI/Fe] $ &    0.007 $\pm$ 0.002 & $-$0.009 $\pm$ 0.011 \\ 
$\rm [TiII/Fe]$ &    0.007 $\pm$ 0.003 &    0.012 $\pm$ 0.014 \\ 
$\rm [V/Fe]   $ & $-$0.013 $\pm$ 0.004 &    0.064 $\pm$ 0.022 \\ 
$\rm [Cr/Fe]  $ &    0.001 $\pm$ 0.002 &    0.005 $\pm$ 0.010 \\ 
$\rm [Mn/Fe]  $ & $-$0.012 $\pm$ 0.003 &    0.038 $\pm$ 0.015 \\ 
$\rm [Co/Fe]  $ &    0.007 $\pm$ 0.003 & $-$0.040 $\pm$ 0.014 \\ 
$\rm [Ni/Fe]  $ &    0.007 $\pm$ 0.003 & $-$0.027 $\pm$ 0.015 \\ 
$\rm [Zn/Fe]  $ &    0.027 $\pm$ 0.006 & $-$0.149 $\pm$ 0.034 \\ 
$\rm [Y/Fe]   $ & $-$0.007 $\pm$ 0.005 &    0.069 $\pm$ 0.025 \\ 
 \hline
\end{tabular}
\end{minipage}
\end{table}

Using these fits, we can obtain GCE corrected abundance ratios: [X/Fe]$_{\rm GCE~corrected}$ = [X/Fe]$_{\rm raw}$ $-$ ($a$ $\times$ Age + $b$) where $a$ and $b$ are from Table \ref{tab:gce_age}. We derived GCE corrected abundances for all objects. In the following discussion and analysis, we adopt these GCE corrected [X/Fe] abundance ratios.

\begin{figure}
\centering
\includegraphics[width=.99\hsize]{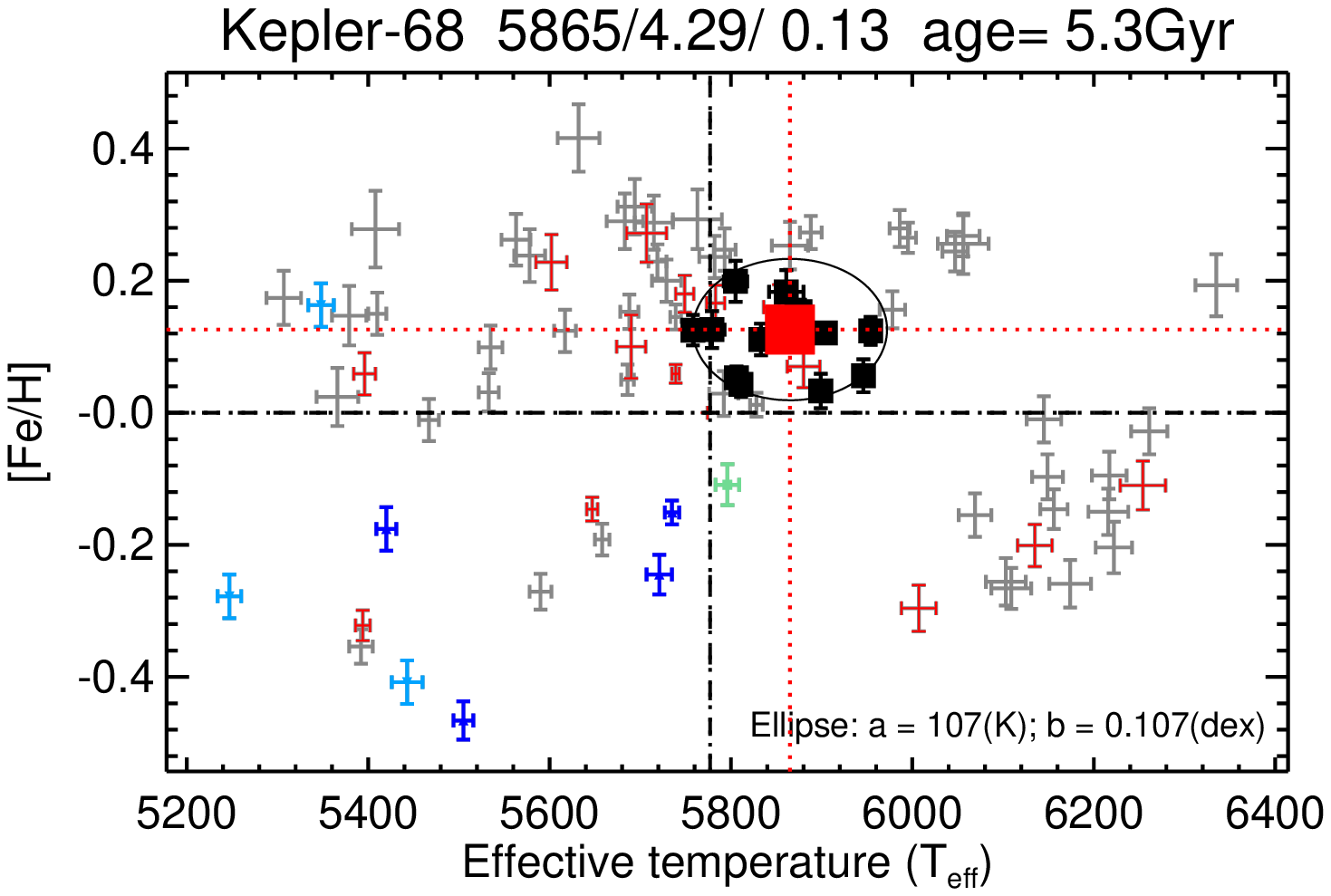}
\caption{[Fe/H] versus effective temperature (\teff). For the planet host Kepler-68 (large red square), the 12 nearest comparison stars are shown as black squares along with their maximum distance in \teff\ and [Fe/H]. An identical approach is applied to obtain the nearest comparison stars for all planet hosts. The colours are as in Figure \ref{fig:xfe_feh_age_a}. 
\label{fig:select}}
\end{figure}

In order to explore the chemical peculiarity of each planet hosting star, we first need to identify the best set of comparison stars by selecting its nearest neighbours in stellar parameter space using the following approach. Taking the planet host Kepler-68 as an example, as shown in Figure \ref{fig:select}, we plot [Fe/H] versus \teff. We define an ellipse with a major axis of 100 K in \teff\ and a minor axis of 0.1 dex in [Fe/H]. Keeping the major to minor axis ratio fixed, we then increase or decrease the ellipse until there are 12 comparison stars within that ellipse (as before, we exclude thick disc, old disc and $s$-rich objects when selecting the nearest comparison stars). In the case of Kepler-68 (Figure \ref{fig:select}), the maximum difference in \teff\ and [Fe/H] between the planet host and the comparison stars was 107 K and 0.107 dex, respectively. When selecting these nearest comparison stars, we arbitrarily imposed a maximum distance of 300 K in \teff\ and 0.3 dex in [Fe/H]. In some instances this resulted in fewer than 12 comparison stars for a given planet host; Kepler-21, Kepler-37, Kepler-408 and EPIC220709978 had 11, 5, 11 and 9 comparison stars, respectively. The rationale in imposing a maximum distance in \teff\ and [Fe/H] for the comparison stars was to ensure that the stellar parameters of the planet hosts and comparison stars were as similar as possible and thereby obtain precise and reliable abundance differences.

We note that our sample spans over a range of [Fe/H], which might induce subtle scatter for our GCE corrections since the slopes of [X/Fe] versus ages can change with [Fe/H]. However, our sample, i.e., each set of a planet host and its comparison stars, is not large enough to be separated into different [Fe/H] bins to minimize such effects. In addition, the age-[Fe/H] relation is basically flat with considerable scatter (see e.g., \citealp{Bensby:2014aa}). For future analysis of a larger sample, it would be useful for GCE corrections to include both age and [Fe/H].

\begin{figure}
\centering
\includegraphics[width=.99\hsize]{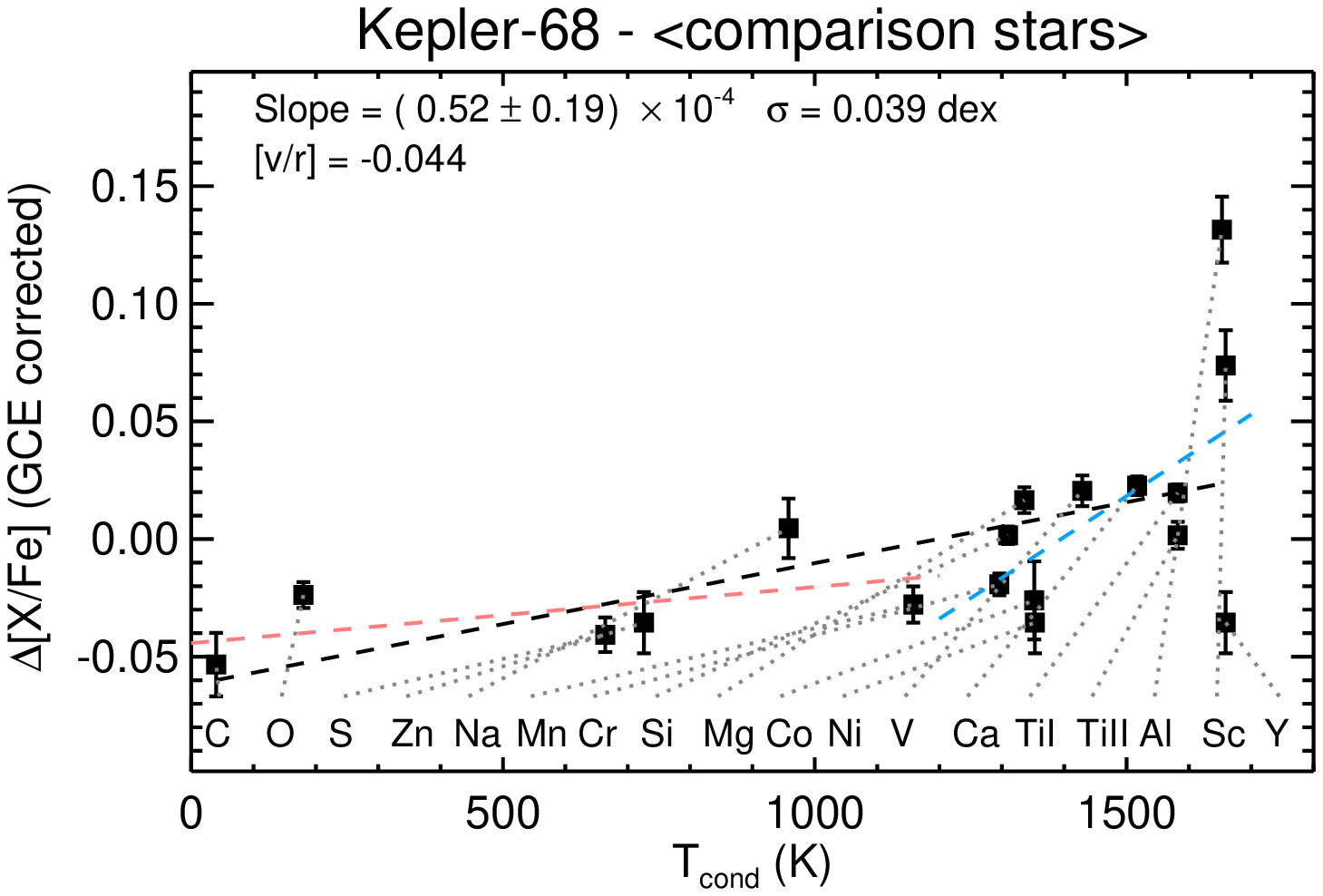} 
\includegraphics[width=.99\hsize]{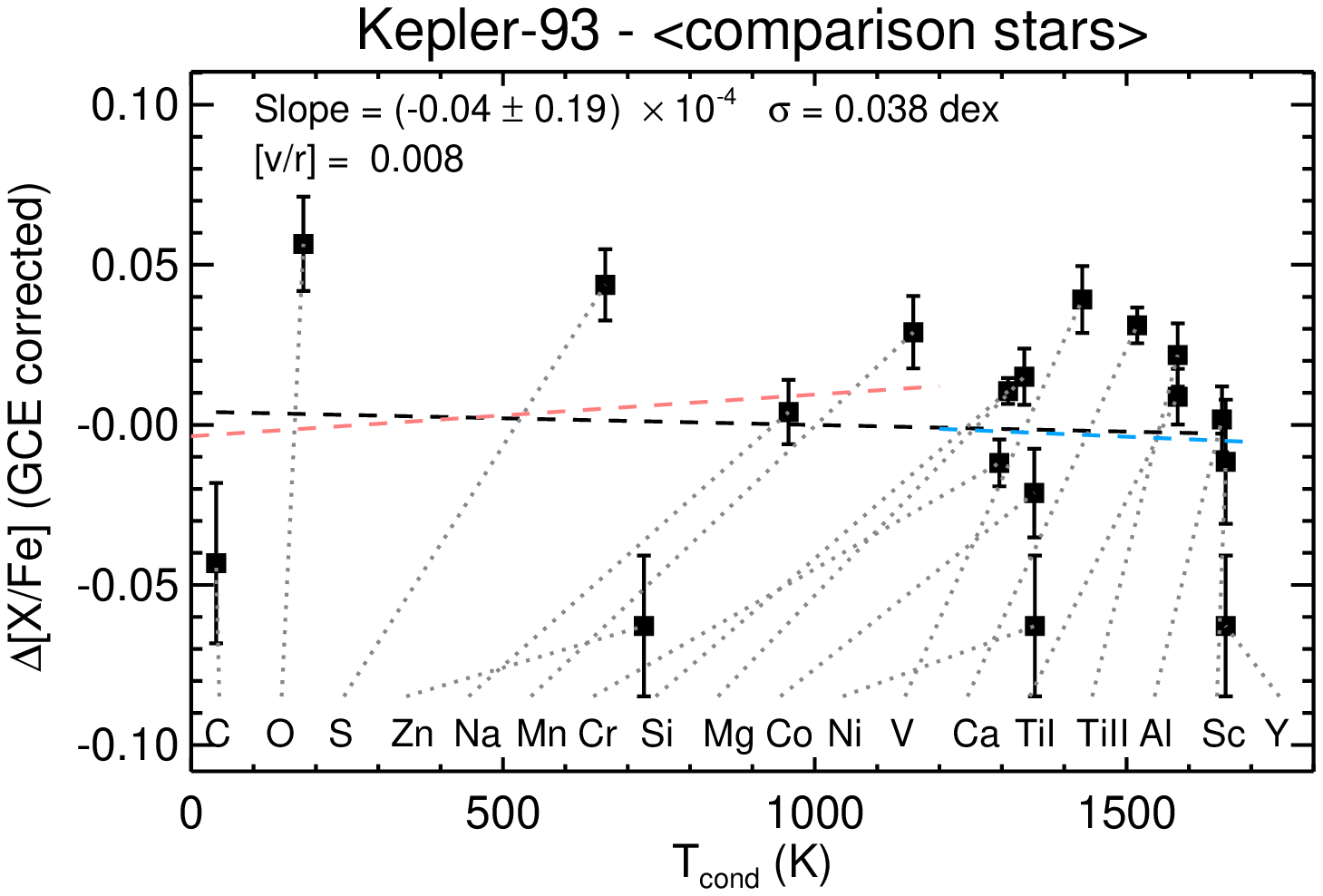}
\includegraphics[width=.99\hsize]{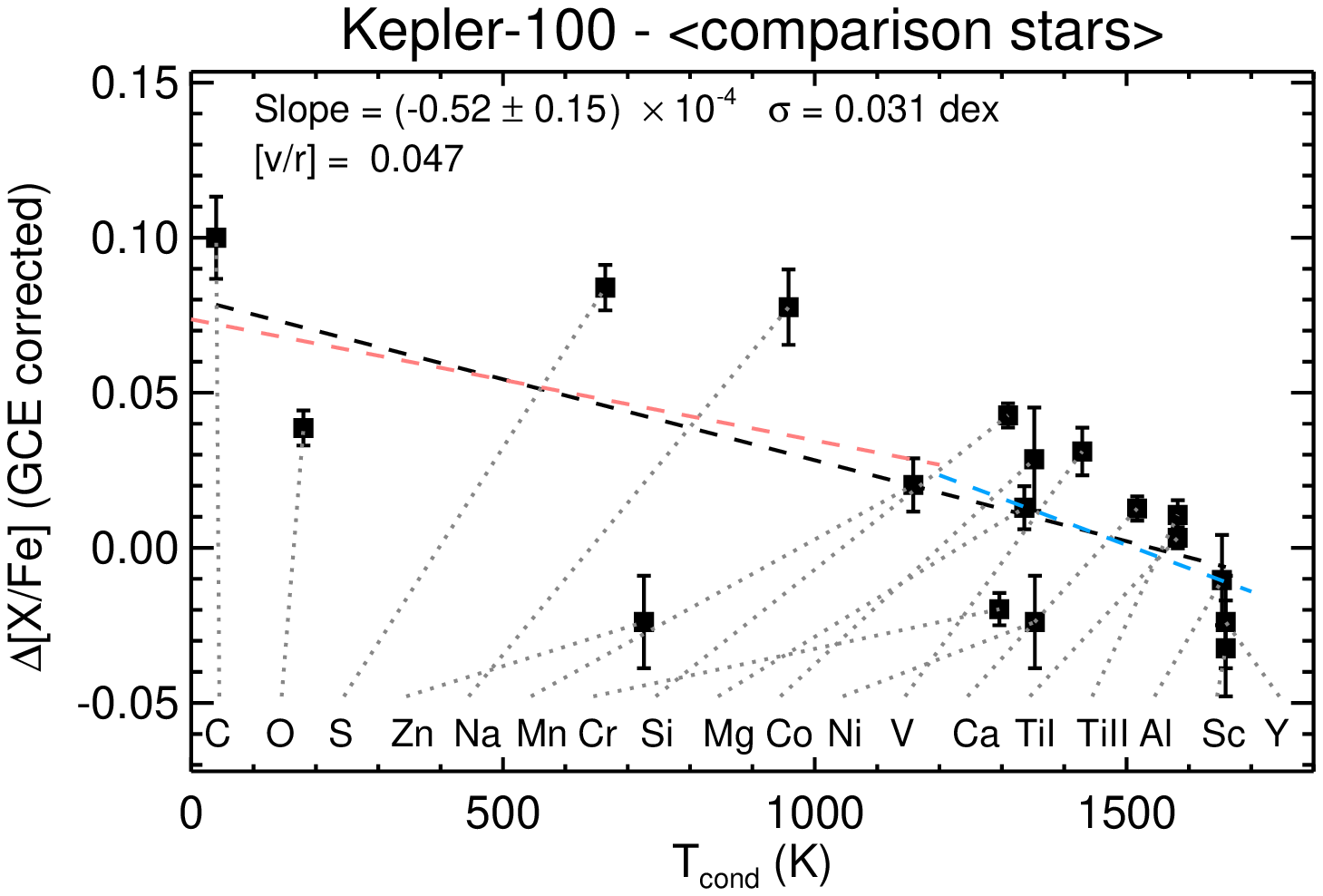}
\caption{Abundance differences between the planet hosts Kepler-68 (upper), Kepler-93 (middle) and Kepler-100 (lower) and the average of the comparison stars ($\Delta$[X/Fe], with GCE corrections applied), versus condensation temperature (\tcond). The error bars represent the total uncertainties as described in the text. The slope, uncertainty and dispersion about the linear fit is shown for all elements (dashed black line). The ratio of volatile (vol; \tcond\ $<$ 1200K) to refractory (ref; \tcond\ $>$ 1200K) elements, [v/r], is also written. The dashed pink and blue lines are the linear fits to the volatile and refractory elements, respectively. 
\label{fig:xfe_tcond}}
\end{figure}

\begin{figure}
\centering
\includegraphics[width=.99\hsize]{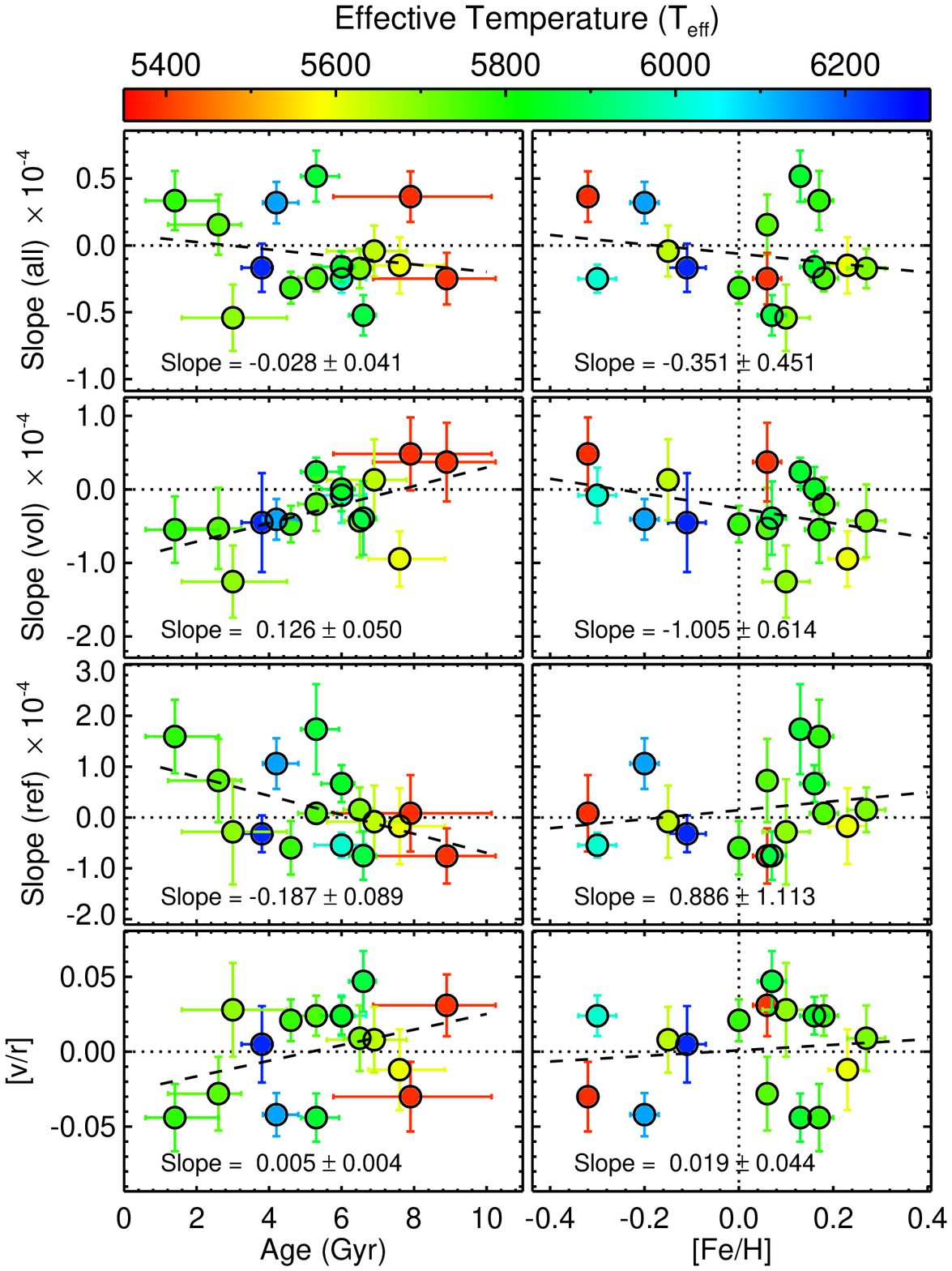}
\caption{Summary of the slope of $\Delta$[X/Fe], GCE corrections applied, versus \tcond\ for the planet hosts versus age (left panels) and [Fe/H] (right panels). The upper panels include the slope of $\Delta$[X/Fe] for all elements. The second top panels include the slope of $\Delta$[X/Fe] for the volatile elements (\tcond\ $<$1200K). The second bottom panels include the slope of $\Delta$[X/Fe] for the refractory elements (\tcond\ $>$1200K). The bottom panels show the average $\Delta$[X/Fe] for the volatile elements minus the average $\Delta$[X/Fe] for the refractory elements. The data are colour coded by the effective temperature. 
\label{fig:tcondsummary} }
\end{figure}

In Figure \ref{fig:xfe_tcond}, we plot the abundance differences (GCE corrected) between Kepler-68 (upper), Kepler-93 (middle) and Kepler-100 (lower) and the average of the 12 nearest comparison stars ($\Delta$[X/Fe]$_{\rm GCE~corrected}$) versus the condensation temperature of each element, \tcond\ taken from \citet{Lodders:2003aa}. The standard errors of the average abundances of comparison stars were taken into account by adding in quadrature to the errors of abundances of the corresponding planet host as the total uncertainties in the abundance differences between each planet host and the average of its comparison stars. These three representative objects exhibit a positive, zero and negative abundance trend with \tcond. In that figure, we highlight four quantities of interest: (1) The slope (and uncertainty and dispersion about the linear fit) for all elements. (2) Same as (1) but for the volatile elements with \tcond\ $<$ 1200 K. (3) Same as (1) but for the refractory elements with \tcond\ $>$ 1200 K. (4) The difference between the average $\Delta$[X/Fe]$_{\rm GCE~corrected}$ for the volatile and refractory elements, [v/r]. We repeated the same exercise for all 16 planet hosts and summarise those analyses in Table \ref{tab:tcond} and in Figure \ref{fig:tcondsummary} where the data are colour coded by the effective temperature. 

\begin{table*}
 \centering
 \begin{minipage}{190mm}
  \caption{Results of the linear fitting of [X/Fe] versus \tcond\ for each planet host. The coefficients are [X/Fe] = $a$ $\times$ \tcond\ + $b$.}
  \label{tab:tcond} 
  \begin{tabular}{@{}lrrrrrrc@{}}
  \hline
Name &
$a$ (all) & 
$b$ (all) & 
$a$ (vol) & 
$b$ (vol) & 
$a$ (ref) & 
$b$ (ref) &
\ldots \\ 
&
(10$^{-4}$ K dex$^{-1}$) &  
(dex) & 
(10$^{-4}$ K dex$^{-1}$) &  
(dex) & 
(10$^{-4}$ K dex$^{-1}$) &  
(dex) \\ 
 \hline 
   \hline
Sun (Iris)      & $-$0.317 $\pm$ 0.119 &    0.038 $\pm$ 0.015 & $-$0.473 $\pm$ 0.248 &    0.044 $\pm$ 0.018 & $-$0.597 $\pm$ 0.526 &    0.038 $\pm$ 0.015 & \ldots \\ 
HD 1461        & $-$0.245 $\pm$ 0.098 &    0.040 $\pm$ 0.013 & $-$0.200 $\pm$ 0.364 &    0.040 $\pm$ 0.027 &    0.082 $\pm$ 0.203 &    0.040 $\pm$ 0.013 & \ldots \\ 
EPIC220709978  & $-$0.249 $\pm$ 0.105 &    0.020 $\pm$ 0.013 & $-$0.080 $\pm$ 0.376 &    0.011 $\pm$ 0.028 & $-$0.545 $\pm$ 0.245 &    0.020 $\pm$ 0.013 & \ldots \\ 
EPIC212357477  & $-$0.541 $\pm$ 0.248 &    0.096 $\pm$ 0.032 & $-$1.254 $\pm$ 0.491 &    0.128 $\pm$ 0.036 & $-$0.282 $\pm$ 1.036 &    0.096 $\pm$ 0.032 & \ldots \\ 
Kepler-37      &    0.365 $\pm$ 0.189 & $-$0.047 $\pm$ 0.025 &    0.483 $\pm$ 0.496 & $-$0.053 $\pm$ 0.037 &    0.083 $\pm$ 0.751 & $-$0.047 $\pm$ 0.025 & \ldots \\ 
Kepler-408     &    0.320 $\pm$ 0.155 & $-$0.048 $\pm$ 0.020 & $-$0.407 $\pm$ 0.276 & $-$0.013 $\pm$ 0.020 &    1.060 $\pm$ 0.498 & $-$0.048 $\pm$ 0.020 & \ldots \\ 
Kepler-21      & $-$0.167 $\pm$ 0.180 &    0.011 $\pm$ 0.023 & $-$0.451 $\pm$ 0.673 &    0.023 $\pm$ 0.050 & $-$0.321 $\pm$ 0.360 &    0.011 $\pm$ 0.023 & \ldots \\ 
HD 179079      & $-$0.172 $\pm$ 0.147 &    0.023 $\pm$ 0.019 & $-$0.428 $\pm$ 0.495 &    0.035 $\pm$ 0.037 &    0.151 $\pm$ 0.439 &    0.023 $\pm$ 0.019 & \ldots \\ 
Kepler-131     &    0.335 $\pm$ 0.221 & $-$0.035 $\pm$ 0.028 & $-$0.548 $\pm$ 0.452 &    0.010 $\pm$ 0.033 &    1.592 $\pm$ 0.723 & $-$0.035 $\pm$ 0.028 & \ldots \\ 
Kepler-68      &    0.518 $\pm$ 0.191 & $-$0.062 $\pm$ 0.025 &    0.239 $\pm$ 0.191 & $-$0.044 $\pm$ 0.014 &    1.736 $\pm$ 0.885 & $-$0.062 $\pm$ 0.025 & \ldots \\ 
Kepler-93      & $-$0.042 $\pm$ 0.190 &    0.004 $\pm$ 0.024 &    0.130 $\pm$ 0.551 & $-$0.004 $\pm$ 0.041 & $-$0.082 $\pm$ 0.711 &    0.004 $\pm$ 0.024 & \ldots \\ 
Kepler-100     & $-$0.522 $\pm$ 0.151 &    0.080 $\pm$ 0.019 & $-$0.391 $\pm$ 0.499 &    0.074 $\pm$ 0.037 & $-$0.750 $\pm$ 0.477 &    0.080 $\pm$ 0.019 & \ldots \\ 
Kepler-409     & $-$0.248 $\pm$ 0.193 &    0.027 $\pm$ 0.025 &    0.372 $\pm$ 0.535 & $-$0.008 $\pm$ 0.043 & $-$0.757 $\pm$ 0.542 &    0.027 $\pm$ 0.025 & \ldots \\ 
Kepler-96      &    0.155 $\pm$ 0.225 & $-$0.029 $\pm$ 0.029 & $-$0.531 $\pm$ 0.552 &    0.003 $\pm$ 0.041 &    0.726 $\pm$ 0.818 & $-$0.029 $\pm$ 0.029 & \ldots \\ 
HD 190360      & $-$0.149 $\pm$ 0.210 &    0.053 $\pm$ 0.028 & $-$0.945 $\pm$ 0.374 &    0.084 $\pm$ 0.028 & $-$0.170 $\pm$ 0.747 &    0.053 $\pm$ 0.028 & \ldots \\ 
HD 219828      & $-$0.160 $\pm$ 0.115 &    0.029 $\pm$ 0.015 &    0.004 $\pm$ 0.300 &    0.026 $\pm$ 0.022 &    0.666 $\pm$ 0.362 &    0.029 $\pm$ 0.015 & \ldots \\ 
 \hline
\end{tabular}
\end{minipage}
This table is published in its entirety in the electronic edition of the paper. A portion is shown here for guidance regarding its content. 
\end{table*}

We also considered only the 10 nearest neighbours (rather than 12) and our conclusions are unchanged. Similarly, our conclusions are unchanged when using the raw [X/Fe] ratios. Finally, we chose not to use the high-fidelity GCE corrections from \citet{Spina:2018mnras} and \citet{Bedell:2018aa} since they were derived from their analysis of a different sample. Had we applied their GCE corrections, however, our conclusions would remain the same. Therefore, we are confident that our results are reliable. 

An independent analysis of a subset of these stars was conducted by \citet{Acuna2019}. There is an excellent agreement for the stellar parameters and abundances between the two analyses. We now briefly discuss the comparison between our results and select literature values (we do not undertake an exhaustive search of the literature for all program stars). 

There are four planet hosts in common between this study and \citet{Schuler:2015aa}; Kepler-21, Kepler-37, Kepler-68 and Kepler-100. Their analysis was based on high-resolution, high $S/N$ spectra. Our stellar parameters are in excellent agreement with \citet{Schuler:2015aa} with average differences (This study $-$ \citeauthor{Schuler:2015aa}) in \teff, \logg\ and [Fe/H] of 10 $\pm$ 23 K, $-$0.01 $\pm$ 0.06 [cgs] and $-$0.01 $\pm$ 0.01 dex, respectively. For the elements common to both studies, we find an average difference (This study $-$ \citeauthor{Schuler:2015aa}) in abundance ratios [X/H] of 0.00 $\pm$ 0.02, 0.02 $\pm$ 0.01, 0.00 $\pm$ 0.01 and 0.00 $\pm$ 0.01 for Kepler-21, Kepler-37, Kepler-68 and Kepler-100, respectively. 

\citet{2017AJ....154..107P} determined stellar parameters for 1305 Kepler objects including nine in common with this study. Our stellar parameters are in good agreement with theirs with average differences (This study $-$ \citeauthor{2017AJ....154..107P}) in \teff, \logg\ and [Fe/H] of 29 $\pm$ 11 K, $-$0.022 $\pm$ 0.023 [cgs] and $-$0.046 $\pm$ 0.007 dex, respectively. 

\citet{2016ApJS..225...32B} measured stellar parameters and chemical abundances for 1617 stars including nine Kepler objects in this study. The agreement in stellar parameters is very good with average differences (This study $-$ \citeauthor{2016ApJS..225...32B}) in \teff, \logg\ and [Fe/H] of 34 $\pm$ 11 K, $-$0.027 $\pm$ 0.022 [cgs] and $-$0.008 $\pm$ 0.014 dex, respectively. For the elements common to both studies, we find an average difference in abundance ratios [X/H] of 0.006 $\pm$ 0.006 dex for the nine common stars. Overall, our results appear to be in very good agreement with literature values.

\section{DISCUSSION} 

As noted in the Introduction, M09 found a striking negative trend between the abundance differences (Sun minus the average for a sample of solar twins) versus \tcond. In Figure \ref{fig:tcondsummary}, the Sun is located at [Fe/H] = 0.00 and has a slope for all elements of $-$0.32 $\pm$ 0.12 $\times$ 10$^{-4}$ dex K$^{-1}$. Among the other 15 planet hosts, nine have negative slopes for the abundance differences versus \tcond\ while the remaining six have positive slopes. There is no significant dependence ($>$ 3-$\sigma$) of these abundance trends as a function of ages or [Fe/H]. Similarly, there is no obvious dependence of these abundance trends with \teff.

The strongest trends are for slope of the ratio of volatile to refractory elements [v/r] versus ages, but they are only significant at the $\sim$ 2-$\sigma$ level. For the Sun, the ratio of volatile to refractory elements is [v/r] = $+$0.021 dex (bottom panel in Figure \ref{fig:tcondsummary}). Among the other 15 planet hosts, nine have positive values of [v/r] while the remaining six have negative values.

Recall that in Figure \ref{fig:tcondsummary}, GCE corrections have been applied. Furthermore, we have also examined similar versions of this plot using the raw [X/Fe] ratios and when applying GCE corrections from \citet{Spina:2018mnras} and \citet{Bedell:2018aa}. In all cases, our conclusions would not change and there are no obvious correlations between abundance trends and ages or [Fe/H].

In the M09 hypothesis, the proto-Sun is surrounded by a proto-planetary disc. In that proto-planetary disc, the refractory elements are preferentially locked-up in planetesimals and subsequently in the terrestrial planets. As a consequence, most of the remaining materials, which are more volatile, in the proto-solar disc are eventually accreted back onto the Sun. The abundance differences between the Sun and the comparison stars could be produced if those comparison stars did not form planets (for whatever reason). Alternatively, planet formation may have proceeded more efficiently in the solar system than for typical stars.

If we assume that the M09 hypothesis is correct, i.e., terrestrial planet formation can induce a subtle chemical abundance signature, then roughly half our sample would appear to exhibit a similar phenomenon (i.e., negative abundance trends versus \tcond) albeit to different degrees. Any scenarios supporting this hypothesis would then need to account for the diversity in abundance differences as a function of \tcond\ amongst the different planet hosting stars.

For the other half of the sample, however, the abundance trends versus \tcond\ exhibit the opposite sign. In the M09 scenario, this could indicate that the comparison stars formed more planets and/or more efficiently than in those planet hosts, or that those planet hosting stars engulfed rocky material through the ingestion of rocky planets, or of giant planets that formed from a sizeable rocky core of a few Earth masses. It is important to recognise that we are implicitly assuming that the comparison stars did not form planets, which might be biased towards the current observations. As mentioned in Section 2, at the time of selection those objects were not known to host planets. However, the occurrence rates of planets in the $Kepler$ field continues to be debated \citep{2019AJ....158..109H} and could be as high as 1.0 \citep{2016RPPh...79c6901B}. Results from microlensing \citep{2012Natur.481..167C} also indicate a high probability that a given star hosts a planet. A high planetary occurrence rate greatly complicates the selection of comparison stars and therefore the identification of abundance differences between planet and non-planet hosting stars. We warn our readers that although the comparison stars do not have planets detected so far, most of them probably have planets. If so, the observed abundance patterns of planet hosts would imply, instead of planet existence effects, the signatures related to the extent of the fractionation of refractory materials from volatile materials during planet formation \citep{Wang:2019icarus}, or to the frequency of planet engulfment events. However, a detailed discussion on such a scenario is beyond the scope of this paper.

The results in this study confirm the considerable diversity in the chemical compositions of planet hosting stars with respect to comparison stars, i.e., a large spread of the \tcond\ slopes of the refractory elements. Interestingly, the total spread in the GCE corrected \tcond\ slopes, $\sim$ 2.5 $\times$ 10$^{-4}$ dex\,K$^{-1}$, is about the same in our sample and in \citet{Bedell:2018aa}. Assuming a range of 800 K for \tcond\ of the refractories, this might indicate that the 'planet-induced effect' has a maximum amplitude of 0.2 dex. This is exactly what has been observed in the most extreme cases of abundance difference in binary pairs: \citet{Ramirez:2019mnras} and \citet{Nagar:2019apjl} found a difference between the binary components HIP 34407/HIP 34426 of 0.2 dex in the refractory elements, and \citet{Oh:2018apj} also found a difference of 0.2 dex in refractory elements for the binary HD 240430 / HD240429. There are a variety of possibilities to explain such chemical diversity which we consider below.

In the context of the M09 hypothesis, the various aspects that can influence the composition of the planet hosting star include the following: (1) The chemical composition of the accreted material: the degree of the change in the composition of the host star will increase with the difference in the composition between the accreted material and the host star. The formation of terrestrial planets, and their numbers, masses (or more explicitly, densities) may induce changes in the planet host's chemical composition. (2) The amount of accreted material: the extent to which the host star's chemical composition changes will depend on the amount of material that is accreted (assuming it is of a different chemical composition). This quantity may be affected, again, by the number of planets formed and their masses/densities. (3) The simulation results by \citet{Bitsch:2018mnras} predicted that different formation locations of planets (inside/outside e.g., H$_2$0 or CO ice line) may play an important role in changing the surface abundance ratios of their host stars for e.g., [Fe/O] or [Fe/C]. (4) The timescale for accretion: during the proto-stellar evolution, the size of the convective envelope decreases such that for a fixed amount of accreted material, a larger convection zone will dilute the accreted material and minimise any potential change in the stellar surface composition. In addition, cooler stars have deeper convection zones, therefore the polluting material should be more diluted in the photosphere of the star. However, we do not find any clear relation between \tcond\ slopes and \teff\ in our study. The lack of correlation between \tcond\ slopes and \teff\ is also reported in \citet{2014A&A...572A..48R}.

Furthermore, recently the M09 signature has been related to the formation of distant giant planets \citep{Booth:2020}, as those planets could trap significant amounts of dust exterior to their orbits, preventing the dust from being accreted into the star, thus explaining the lower abundance of refractories in the Sun.

Another potential scenario, namely planet engulfment scenario, was proposed initially by \citet{Pinsonneault:2001apjl} to explain the differences in elemental abundances in stars with and without planets. In this scenario, the chemical compositions of the planet-hosting star might be enhanced in refractory elements because of the post-formation accretion of inner planets/material. This scenario is based on the assumption that the presence of a close-in gaseous planet might scatter or otherwise perturb the inner planets onto the surface of their host star, thus adding a large amount of H-depleted material, enhancing the stellar photospheric abundances for refractories. This has been reported and discussed in several spectroscopic studies of wide binaries by e.g., \citet{Oh:2018apj}, \citet{Liu:2018aa}, \citet{Ramirez:2019mnras} and \citet{Nagar:2019apjl}, where a planet-hosting component in a binary system could be enriched in refractories, when compared to another component in this system. Similar phenomena have been found in open clusters, where the peculiar star in on open cluster is more metal-rich than the average of the cluster stars (e.g., \citealp{Spina:2015aa}). In addition to the planet formation scenario proposed by M09, the planet engulfment scenario can also explain the mixed observed \tcond\ trends in our study. For example, stellar-planetary systems with close-in Neptunes or Jupiters should have experienced strong dynamical interactions in the past, thus it is possible that they have induced more material to fall into the star. It is possible that those planet hosts showing positive \tcond\ trends have engulfed more material from their planetary systems depending on the architecture of the system.

Finally, there are alternative possibilities to explain the diversity in the chemical compositions of planet hosting stars that do not involve planets (e.g., \citealt{2011A&A...528A..85O}). Inhomogeneous chemical evolution would produce element-to-element differences amongst stars. In that scenario, one might expect to see chemical abundance differences possibly attributed to different nucleosynthetic processes. While we have attempted to correct for Galactic chemical evolution, our sample is small and the corrections applied may not be sufficiently accurate to reveal subtle abundance trends amongst our sample. In addition, the \tcond\ of \citet{Lodders:2003aa} is calculated based on the solar elemental abundances at a fixed solar nebular pressure. However, studies have shown that the sequence of condensation and thus the composition of condensed materials would vary depending on the nebular pressures and the stellar elemental abundances \citep{Ebel:2000GeCoA,Unterborn:2017apj}. This may further complicate the exploration of planet signatures based on the abundance versus \tcond\ trends.

\section{CONCLUSIONS} 

We present a differential analysis of 16 planet hosting stars and 68 comparison objects using high resolution, high $S/N$ spectra. We obtained high-precision stellar parameters and relative chemical abundance ratios. Average uncertainties for \teff, \logg\ and [Fe/H] are 15 K, 0.034 [cgs], 0.012 dex, respectively. The average uncertainties for abundance ratios [X/H] range from 0.010 dex (Si) to 0.042 dex (C). The abundance ratios are corrected for GCE using the trends between [X/Fe] versus ages. For each planet host, we identify the nearest comparison stars in terms of effective temperature and metallicity. We then examine the abundance differences between the planet host and the average of the comparison stars as a function of dust condensation temperature. We confirm that the Sun is chemically different from its comparison stars; the trend between abundance differences and \tcond\ is negative. While about half of the planet hosts exhibit similar negative \tcond\ $-$ abundance trends, the other half, however, exhibit positive \tcond\ $-$ abundance trends. 

It remains unclear why there is such chemical diversity among the planet hosting stars with respect to the comparison stars. The range of \tcond\ slopes observed might reflect the range of possible planet-induced effects present in the planet hosting stars, from the sequestration of rocky material (refractory poor), to the possible ingestion of planets (refractory rich). In the context of the M09 work and potential chemical signatures of planet formation, the timescale, location, efficiency and degree of planet formation could all affect stellar chemical compositions. Indeed, there are considerable challenges in finding an appropriate comparison sample for each of the planet hosts. Alternatively, planet engulfment, inhomogeneous chemical evolution, or inconsistent condensation processes amongst different stars, could also explain such differences.

Finally, following \citet{2019MNRAS.482.2222W}, in a future paper (Wang et al.\ in prep) we will use the high-precision chemical abundances of planet hosts from this paper to infer the bulk composition and internal structure of exoplanets.

\section*{Acknowledgments}

The authors wish to recognize and acknowledge the very significant cultural role and reverence that the summit of Maunakea has always had within the indigenous Hawaiian community. We are most fortunate to have the opportunity to conduct observations from this mountain. Australian community access to the Keck Observatory was supported through the Australian Government's National Collaborative Research Infrastructure Strategy, via the Department of Education and Training, and an Australian Government astronomy research infrastructure grant, via the Department of Industry and Science.
This work has made use of data from the European Space Agency (ESA) mission {\it Gaia} (\url{https://www.cosmos.esa.int/gaia}), processed by the {\it Gaia} Data Processing and Analysis Consortium (DPAC, \url{https://www.cosmos.esa.int/web/gaia/dpac/consortium}). Funding for the DPAC has been provided by national institutions, in particular the institutions participating in the {\it Gaia} Multilateral Agreement. 
We gratefully acknowledge support from the Australian Research Council (grants FL110100012, FT140100554, and DP120100991). F.L. acknowledges support from the grant "The New Milky Way" from the Knut and Alice Wallenberg Foundation and the grant 184/14 from the Swedish National Space Agency. Contributions of H.S.W. were carried out within the framework of the NCCR PlanetS (Project 4.2) supported by the Swiss National Science Foundation. L.S. acknowledges financial support from the Australian Research Council (Discovery Project 170100521) and from the Australian Research Council Centre of Excellence for All Sky Astrophysics in 3 Dimensions (ASTRO 3D), through project number CE170100013. J.M. thanks support from FAPESP (2018/04055-8) and CNPq (Bolsa de Produtividade).
F.L. acknowledges suggestions from Prof. Michael Murphy during manuscript writing. At last, we thank the referee for the valuable comments that helped to improve our manuscript.


\appendix
\section{Extra figure}

\begin{figure}
\centering
\includegraphics[width=.99\hsize]{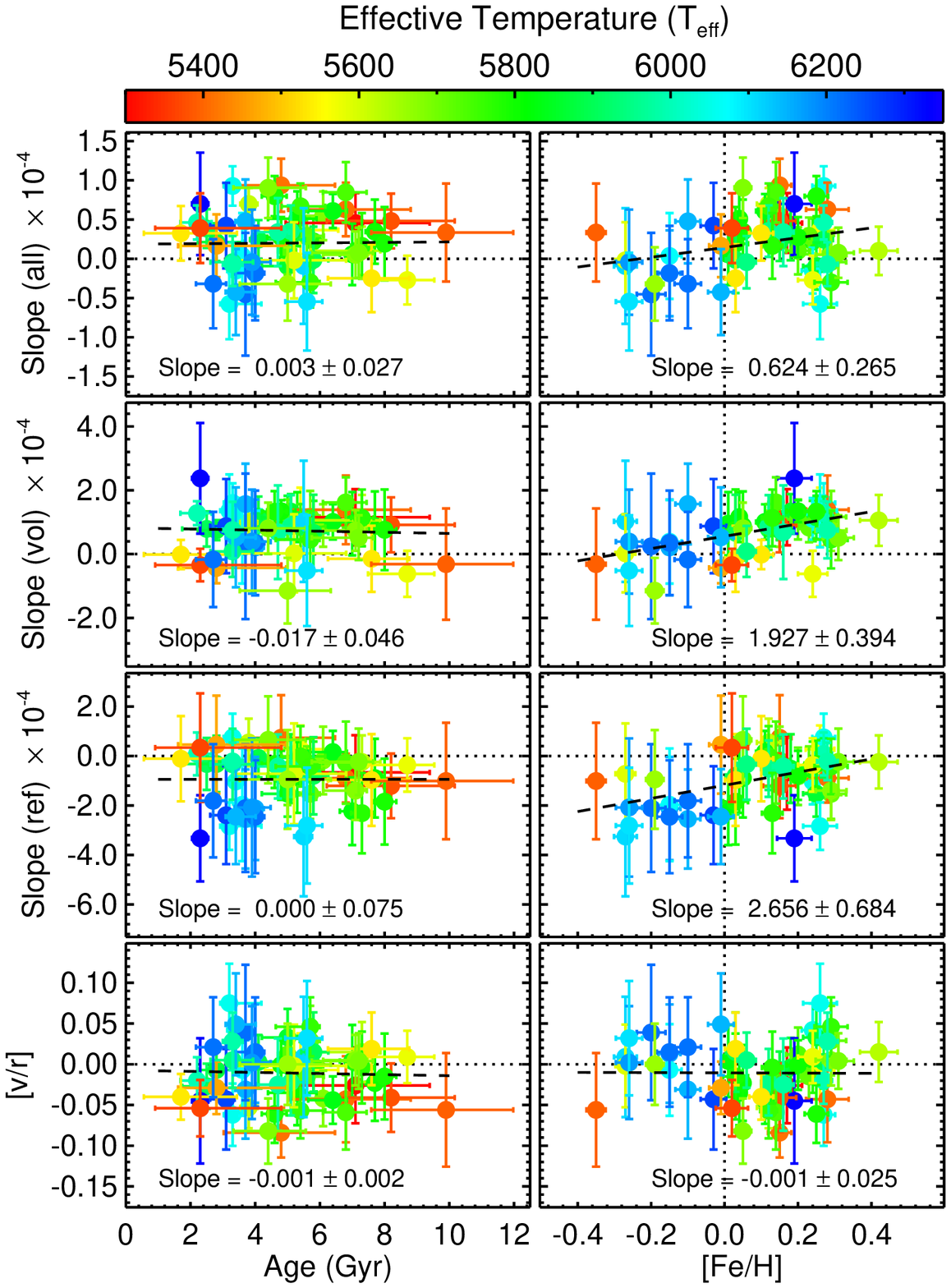}
\caption{Same as Figure \ref{fig:tcondsummary} but for the comparison stars.
\label{fig:tcondsummary2} }
\end{figure}

In Figure \ref{fig:tcondsummary2}, we show the same summary as in Figure \ref{fig:tcondsummary} but for the comparison stars. There are subtle trends with [Fe/H], which would suggest that it is important to restrict the range in [Fe/H] for the comparison stars. 

\section*{SUPPLEMENTARY MATERIAL}

The following supplementary material is available for this
article online:

Table A1. The line list and equivalent width measurements adopted for our analysis.

Full version of Table 1, 2, and 5.

\label{lastpage}

\end{document}